# On the origin of in-gap states in homogeneously disordered ultrathin films. MoC case.

V. Hašková,[1] M. Kopčík,[1] P. Szabó,[1,*] T. Samuely,[1] J. Kačmarčík,[1] O. Onufriienko,[1] M. Žemlička,[2] P. Neilinger,[2] M. Grajcar[2] and P. Samuely[1,*]

[1]Centre of Low Temperature Physics, Institute of Experimental Physics, Slovak Academy of Sciences and P. J. Šafárik University, SK-04001 Košice, Slovakia
[2]Department of Experimental Physics, Comenius University, SK-84248 Bratislava, Slovakia

Many disordered superconducting films exhibit smeared tunneling spectra with evident in-gap states. We demonstrated that the tunneling density of states in ultrathin MoC films is gapless and can be described by the Dynes version of the BCS density of states with a strong broadening parameter Γ accounting for the suppression of coherence peaks and increased in-gap states. The thinner the film, the lower the $T_c$ and the superconducting energy gap Δ and the larger the Γ. MoC films of 3 nm thickness deposited simultaneously on silicon and sapphire substrates reveal very similar scalar disorder, evidenced by the equal sheet resistance, but exhibit different superconducting characteristics of $T_c$, Δ and Γ, suggesting that pair breaking responsible for the dissipation channel and the suppression of superconductivity originates on the film-substrate interface. It indicates that sapphire is a stronger pair breaker. Interface pair breaking can be operative in other cases as well.

## I. INTRODUCTION

By combining the state-of-the-art techniques of thin film deposition and scanning tunneling microscopy (STM) one can study the evolution of the physical properties of ultra-thin films as their thickness reaches the nanoscale region, where the thin film-substrate interface can play a significant role. Moreover, the superconducting state is highly volatile in the presence of any pair breaking effects. Therefore, if the interface with the substrate acts as a local source of magnetism, a superconducting ultra-thin film becomes a sensitive probe suitable for studying the substrate influence.

The local STM measurements of the superconducting density of states allow direct study of the homogeneity of the superconducting phase and thus the presence of various pair breaking effects can be detected [1-5]. The density of states of standard BCS s-wave superconductors is characterized by a gap and coherence peaks at the gap edges. Anderson proved [6] that even in dirty superconductors with the electron mean free path lower than the coherence length, the pairing of time-reversed degenerate states led to unchanged superconducting energy gap. On the other hand, Abrikosov and Gor'kov (AG) have shown [1] that magnetic impurities breaking the time reversal symmetry leads to a depression of $T_c$ and strong modification of BCS density of states (DOS). Maki [2] and de Gennes [3] noticed that the AG results are applicable on many different pair-breaking perturbations such as magnetic fields, current, spin exchange, spatial gradients of the order parameter and so on. Skalski et al. [4] have calculated the DOS and showed that in the presence of pair breaking characterized by parameter α (α = ℏ/2τ, τ is quasiparticle lifetime) the minimum excitation energy $E_G$ is not equal to the gap-parameter Δ (which is a measure of the pairing strength). The increase of α broadens the coherence peaks and $E_G$ shrinks to lower values until it vanishes at α/Δ ≈ 1 when the DOS becomes gapless.

In our previous scanning tunneling microscopy and spectroscopy (STM/STS) experiments on homogeneously disordered MoC ultrathin films we observed much stronger broadening of DOS which in fact leads to a gapless quasiparticle spectrum already for the samples characterized by $T_c$ only slightly lower than in the bulk and ordered material. The disorder level of the samples was characterized by the $k_F l$ product, where $l$ is the electron mean free path and $k_F$ is the Fermi momentum. The thinnest samples had $k_F l$ close to unity. The tunneling spectra could be described by the phenomenological Dynes formula [7] introducing a broadening parameter Γ to BCS DOS, where the suppression of coherence peaks is accompanied by an increase of the quasiparticle states inside the ideal gap. Upon reducing the film thickness, as $k_F l$ decreases, the superconducting energy gap is reduced at the same rate as the transition temperature, also the coherence peaks are suppressed and the in-gap states increase [5]. Remarkably, the in-gap states in superconducting DOS spectrum have been detected by STM tunneling experiments also in disordered NbN thin films [8,9]. Neither of these experiments can be described by the Skalski version of the superconducting DOS [5]. The authors used the Dynes formula to fit the smeared spectral features of tunneling spectra. Also, microwave characteristics of disordered films show significant deviations from the standard BCS theory. The imaginary part of the complex conductivity in TiN films [10] could be explained in a narrow temperature window by a Skalski-like DOS but understanding of the real part, which is mostly influenced by disorder, is far from being complete. Anomalous dissipative conductivity has also been found via terahertz spectroscopy on thin films of molybdenum nitride [11]. Our previous investigations of the complex conductivity of highly disordered MoC superconducting films with $k_F l \approx 1$ show high resistive losses in the real part of conductivity [12] which are related to the finite quasiparticle DOS at the Fermi level found by direct STM tunneling measurements on the same film.

The Dynes formula is derived from a purely phenomenological model, however, very recently, *Herman & Hlubina* [13] assigned intrinsic microscopic explanation to it as they derived the Dynes formula for a superconductor with a Lorentzian distribution of local pair-breaking fields and arbitrary potential disorder. They dubbed such systems "Dynes superconductors". This approach might provide a clue to the detected finite DOS at the Fermi level found in many disordered systems for as long as the source of the pair-breaking fields is identified. In the case of ultrathin films, the interface between the film and the substrate breaking translational symmetry allowing for uncompensated electron orbitals can be such a source.

In this paper we show that indeed, the interface strongly influences the superconductivity of the thin films. First, we show that $\Delta$ and $T_c$ are larger on thicker films or local film areas and anticorrelate with the smearing parameter $\Gamma$, i.e. the closer to the interface, the weaker the superconducting state. Second, we prepared 3 nm thin MoC films simultaneously on sapphire and silicon substrates. The films exhibit equal sheet resistance indicating a similar level of scalar disorder affecting the mean free path of the quasiparticles, however, their superconducting properties such as the critical temperature $T_c$, the energy gap $\Delta$ and the broadening parameter $\Gamma$ accounting for the in-gap states and the suppression of coherence peaks are different. We attribute this effect to a different pair breaking at the two interfaces.

## II.  EXPERIMENT

MoC films of 3 nm thickness were prepared by reactive magnetron sputtering in an argon-acetylene mixture [14]. Single atoms ejected from the surface of molybdenum target react with carbon atoms within the acetylene molecules and the product is subsequently deposited on the substrate. The superconducting properties of $MoC_x$ samples strongly depend on the carbon concentration [15]. Varying the acetylene pressure at the sputtering we tuned the carbon concentration and obtained samples with the highest possible $T_c$. The crystallographic structure of MoC thin films was characterized by XRD measurements on 30 nm MoC films. The relative positions of the diffraction peaks are associated with a cubic crystallographic structure of the δ-phase of MoC with the atomic lattice constant $a_0 = 0.42$ nm. This phase has the highest bulk critical temperature from the MoC phase diagram [15]. For our experiment, two types of substrates were used: monocrystalline c-cut sapphire ($Al_2O_3$) and [100] silicon (Si). The Si substrate was not cleaned in-situ before the sputtering. Therefore, in the case of MoC/Si films the MoC phase was deposited to native amorphous $SiO_x$ surface [16]. Both MoC/Si and MoC/$Al_2O_3$ samples were prepared simultaneously in a single sputtering procedure, during which the substrates were kept at 200°C as suggested by *Lee and Ketterson* [17]. To ensure that during a single sputtering process all the samples develop identical MoC stoichiometry a test was performed where only $Al_2O_3$ substrates were placed on all sample holders in the sputtering set up. Subsequent analysis of these films revealed no differences in $R_s$ and $T_c$, verifying identical stoichiometry. The sputtering rate of ~ 9 nm/min was used and the final film thickness was verified by the X-ray reflectometry measurements. Following the sputtering process, the samples were coated with a protective photoresist layer to prevent excessive oxidation of the surface. The coating was dissolved prior to the STM experiments.

Transport measurements of the sheet resistance were performed using the Van der Pauw four-probe technique [18]. The scanning tunneling microscopy and spectroscopy measurements were carried out via a sub-Kelvin STM system developed in the Centre of Low Temperature Physics, allowing experiments down to $T = 400$ mK temperatures and magnetic fields up to $B = 8$ T. The surface topographies were recorded using a gold tip in the STM constant current mode at the tunneling resistance range of 10 – 50 MΩ with bias voltages in the range of 10-100 mV.

## III.  RESULTS

Figure 1 shows typical temperature dependencies of sheet resistance $R_s(T)$ of MoC/Si and MoC/$Al_2O_3$ samples. The main panel of Fig. 1 presents the temperature dependence of $R_s$ from $T = 200$ K down to the superconducting transition showing a weak increase with a negative curvature when approaching $T_c$ for both films. This increase, typical for disordered thin films, can be described within the framework of quantum corrections [19,20] to the standard Drude conductivity of disordered metals (weak localization and electron-electron interactions). The inset of Fig. 1 shows a low temperature zoom of sheet resistance $R_s$ close to the transition temperature. The sheet resistance curves reveal similar temperature dependency in both cases with maximum values at about $R_s \approx 1.1$ kΩ, proving that the disorder affecting the electronic mean free path in these samples is very similar. On the other hand, the samples differ significantly in their transition temperatures, determined at 50% of the resistive transition, which are 3.5 K and 2.5 K for MoC/Si and MoC/$Al_2O_3$ samples, respectively. The sharp single-phase superconducting transition with $\Delta T_c/T_c \approx 0.25$ width, where $\Delta T_c$ is the difference in $T_c$ determined at 90% and 10% of the normal state resistance $R_N = 1120$ Ω, suggest the presence of homogeneous disorder in the films.

The surface of the studied MoC thin films was investigated by means of STM topography measurements. Figure 2 shows typical images of 200 x 200 $nm^2$ surface measured on MoC/$Al_2O_3$ (a) and MoC/Si (b) films at $T =$

450 mK. MoC/Al$_2$O$_3$ films display areas where clean polycrystalline surfaces were observed consisting of single-crystalline grains of the lateral size 20-40 nm. The growth of MoC films is different on the Si substrate. Surface topography observed on MoC/Si films depicted in Fig. 2b) shows morphology with characteristic 3 – 6 nm wide "boomerang-like" structures. We observed atomically resolved surfaces on both samples. The upper insets in Fig. 2 a) and b) show atomic resolution topography of a typical nanocrystalline grain. In both sets of the studied samples we observed distorted hexagonal lattice structure. Fourier transform images of the atomic structures are shown in the bottom insets of Fig. 2a) and b). The Fourier patterns are located in 6 points of a hexagon at 1/0.6 nm$^{-1}$ positions. The large (a = 0.6 nm) lattice parameter indicates that the atoms are located in the most dense 111 plane of a cubic B1 type lattice with the lattice parameter $a = a_0\sqrt{2}$. B1 lattice is typical for the δ-MoC phase [15] with a lattice constant $a_0$ = 0.42 nm, identified by XRD measurements on thicker films. STM measurements on both sets of thin films revealed *rms* roughness of 0.6-0.7 nm.

The local STS measurements were performed in CITS (Current Imaging Tunneling Spectroscopy) mode [21], where *I-V* curves are measured in a grid of 128x128 points in a 100x100 nm$^2$ surface area at *T* = 450 mK and numerically differentiated in order to obtain the tunneling conductance *dI/dV* spectra. The tunneling conductance measured through a tunnel junction between a normal metal and a superconductor (N-I-S) can be described as

$$\frac{dI}{dV} = N_N \int_{-\infty}^{\infty} N_S \left[ \frac{-\partial f(E+eV)}{\partial eV} \right] dE,$$

where *E* is the energy, the term in the square bracket represents thermal smearing and $N_N$ is the density of states (DOS) of normal metal. $N_S$ represents the superconducting DOS, which is defined in the BCS theory as

$$N_S(E) = Re \left( \frac{E}{\sqrt{E^2 - \Delta^2}} \right)$$

where Δ is the superconducting energy gap. Since the DOS of a normal metal is constant at low energies, each tunneling conductance spectrum measured at low temperatures becomes directly proportional to the DOS of the superconductor. In our case the tunneling spectra are strongly broadened and high level of in-gap states is observed even at the lowest measured temperature of 450 mK (Fig. 4) where the thermal smearing should amply allow for vanishing differential conductance. Such data can be described by the Dynes modification of the BCS DOS with a spectral smearing Γ incorporated into the complex energy as $E = E' - i\Gamma$ [7].

The values of the superconducting energy gap Δ and spectral smearing Γ of 128 x 128 spectra measured on a 100 x 100 nm$^2$ surface area are shown in Fig. 3. The gap values (Fig. 3a) are distributed around the modes <$\Delta_{Si}$>= 0.66 meV and <$\Delta_{Al2O3}$>= 0.5 meV with halfwidths $w_{Si}$ = 0.09 meV and $w_{Al2O3}$ = 0.1 meV with standard deviation below 4%. The distribution of the Dynes smearing parameter Γ for both samples determined in the same fitting procedure is shown in Fig. 3b). The smearing parameter for MoC/Si samples is distributed around the mode <$\Gamma_{Si}$>= 0.2 meV, with halfwidth $u_{Si}$ = 0.089 meV. The mode of the MoC/Al$_2$O$_3$ sample is <$\Gamma_{Al2O3}$> = 0.225 meV and $u_{Al2O3}$ = 0.11 meV. In concordance with our former findings [5], we attribute this distribution to the noise of the apparatus and sample morphology (see below).

Typical tunneling conductance spectra measured at *T* = 450 mK in MoC/Al$_2$O$_3$ and MoC/Si films are shown in Fig. 4 with solid lines. The symbols plot the fitting curves with the fitting parameters for MoC/Si: $\Gamma_{Si}$ = 0.17 ± 0.01 meV, $\Delta_{Si}$ = 0.65 ± 0.02 meV - open black circles and MoC/Al$_2$O$_3$: $\Gamma_{Al2O3}$ = 0.237 ± 0.03 meV, $\Delta_{Al2O3}$ = 0.51 ± 0.03 meV - open gray triangles. These fitting parameters coincide well with the modal values determined above by the statistical analysis of the spectral maps, therefore the spectra can be considered characteristic for each thin film. The inset of Fig. 4 shows temperature dependencies of the energy gap of both samples (MoC/Si – circles, MoC/Al$_2$O$_3$ – triangles) determined by fitting the temperature dependencies of the tunneling spectra on the main panel. The locally determined critical temperatures of our samples are: $T_c$ = 4 K for MoC/Si and $T_c$ = 3.2 K for MoC/Al$_2$O$_3$ which corresponds to the onset of superconductivity at 90% of $R_N$ shown in Fig. 1. Hence, from the results described above follows that despite nearly identical level of disorder manifested by the sheet resistance $R_s$, the two films deposited on different substrates exhibit substantially different superconducting properties. On the other hand, the coupling strength $2\Delta/k_BT_c$ is equal to 3.8 not only for our 3 nm samples made on different substrates, but also for all MoC thin films of different thicknesses [5].

We further elaborate on the influence of the substrate by examining the local superconducting properties with respect to the sample morphology. These measurements have been performed in areas with well-defined protruding structures. Figure 5 shows the spatial variation of the fitting parameters Γ and Δ across a 40 x 50 nm$^2$ area of MoC/Si thin film surface. The distinctive boomerang structure protruding by ca. 1.3 nm appears red in topography (Fig. 5a). Fig. 5b) represents the gap-map, which plots the values of Δ, obtained from fitting of the tunneling spectra measured at each point in the area. Fig 5c shows the image of the corresponding Γ values, the gamma-map. By using the same relative color contrast, panel c) of Fig. 5 appears as a negative of panel b). Comparing the gap-map and gamma-map images with the surface topography we can see, that in the red area of the "boomerang" the value of the energy gap Δ is 20% higher (Δ ≈ 0.72 meV), than in the lower flat blue areas (Δ ≈ 0.6 meV). The Γ values show the opposite tendency. Γ is approx. 40% lower at the top of the boomerang (Γ ≈ 0.15 meV) compared to the bottom (Γ ≈ 0.24 meV). In other words, areas closer to the substrate feature lower Δ values and higher Γ values than the surface of the protruding "boomerang". The dependence of the energy gap Δ on the spectral smearing Γ (from Fig. 5b, c) is shown with black symbols in Fig. 5d. The red dashed line is a linear fit to emphasize the linear suppression of Δ with increasing Γ, indicating that the mechanism responsible for the suppression of superconductivity is a pair breaking at the interface.

The effect of surface corrugation on MoC/Al$_2$O$_3$ samples on the tunneling spectra have been addressed in our previous paper [5], where we found a correlation between the size of the superconducting gap and the local thickness of the sample. As can be seen in Fig. 3 the distribution of the gap and the smearing parameter in both types of samples is very similar. On the other hand, the modes differ substantially. This indicates that rather than the local inhomogeneities, the modes are driving the global superconducting properties.

## IV.  DISCUSSION

A thickness dependence of the superconducting characteristics has been observed in MoC samples sputtered onto sapphire substrates, published in our previous paper [5]. There, we have shown that the reduction of the MoC thin film thickness from 30 nm to 3 nm leads to a concurrent decrease of both Δ and $T_c$ (maintaining a constant ratio of $2\Delta/k_BT_c$) and an increase of Γ. However, in that case the change in superconducting parameters was accompanied by the change of the sheet resistance. According to the standard theory [22-24] the effect of reduction of $T_c$ upon increased disorder with decreased thickness of the films is a result of a renormalization of the Coulomb interaction. In accordance with the Finkelstein's model [24] obtained by the renormalization group analysis, the suppression of $T_c$ is directly governed by increasing sheet resistance $R_s$. Indeed, the $T_c(R_s)$ dependence of a series of MoC films with different thickness could be perfectly fitted by the Finkelstein's formula [25]. However, this mechanism is apparently not the only one operative in the case of our 3 nm films prepared on Si and Al$_2$O$_3$ substrates featuring exactly equal sheet resistance but different $T_c$.

Our samples deposited on different substrates differ in $T_c$, Δ and Γ. Namely, the MoC/Al$_2$O$_3$ sample with lower Δ and $T_c$ features a higher Γ, while in the case of MoC/Si higher Δ and $T_c$ are accompanied by lower broadening parameter Γ. A particular interface comes as a natural source of specific pair breaking leading to different superconducting characteristics while the normal properties determined by the scalar disorder inside the films stay the same. Remarkably, in the previous paper [5] a finite Γ parameter appeared on the films with thickness lower than the superconducting coherence length (10, 5 and 3 nm thickness). On the contrary, the tunneling spectrum of the 30 nm MoC film features a hard gap with vanishing broadening parameter Γ in a perfect agreement with the BCS prescription. Since the tunneling spectroscopy is sensitive to the sample area of the order of the coherence length, processes at the interface should be reflected in the STS characteristics of sufficiently thin films.

The differences in the spectral smearing and the level of in-gap states between the samples can be naturally explained within the *Herman & Hlubina* model. Here, the Dynes formula describes superconductors with a specific distribution of local pair-breaking fields and arbitrary potential disorder. In this model the energy gap is reduced with increasing pair breaking effect (quantified by Γ). The question remaining to be answered is the source of pair-breaking fields responsible for Γ parameter at a particular interface. Interdiffusion of oxygen atoms at the substrate-thin film interface can form uncompensated electron orbitals [26]. Notably, a possible source of local magnetism is the presence of structural surface defects on the insulating oxide substrates with unpaired spins. The native spin polarized defects localized in the place of oxygen or cation vacancies are even capable of inducing ferromagnetism [27-30]. The intrinsic magnetism on silicon surface step edges is discussed in [27]. Recent studies on inorganic nanoparticles have shown room-temperature ferromagnetism to be a universal characteristic of these nanomaterials, such a typical material exhibiting ferromagnetism also being Al$_2$O$_3$ [29]. It is noteworthy that Proslier *et al.* [26] also suggested to interpret their broadened tunneling spectra of Nb in terms of magnetic scattering on accidental sub-stoichiometric Nb$_2$O$_5$ oxides producing unscreened *d*-band magnetic moments.

The reduction of the gap Δ with increasing spectral smearing Γ shown in Fig. 5d is a very strong evidence of the pair breaking origin of the suppression of superconductivity [3,4,13]. The Hermann & Hlubina model can describe this suppression only qualitatively. The model is taking into account only the local pair breaking effects. However, the influence of locally varying disorder defined with the Finkelstein's theory is not considered. This might be the reason why our Δ(Γ) values are below the theoretical curve for Dynes superconductors, shown in Fig. 7 of Ref. [13]. Further studies are necessary. Accordingly, the thickness dependence of local Δ and Γ shown in Fig. 5 can be explained by a varying proximity to the pair-breaking fields at the interface. Such a model also corroborates the increase of the Γ parameter found in the MoC films upon decreasing thickness in our previous studies [5], while the increased disorder capable to explain the reduction of $T_c$ and Δ provides no explanation for the appearance of finite Γ.

Accordingly, we attribute different superconducting characteristics of the films to different pair-breaking effects of two interfaces. We emphasize that both samples were prepared simultaneously and sputtering process was homogeneous, thus the possibility of different stoichiometry is out of consideration. Another important issue is that our MoC films belong to the class of homogeneously disordered systems rather than granular disordered films. As discussed in the review article of Gantmakher and Dolgopolov [22], increasing disorder in the former drives the transition temperature to the lower values while the transitions remain sharp and single-step due to transparent grain boundaries. This is the case of our MoC films as shown in Fig. 1 for 3-nm films and also in Ref. 7 for variety of thicknesses/disorder/sheet resistances. Overall, this behavior is clearly distinct from the case of granular superconductivity, where the grain boundaries form tunneling barriers and the sample becomes an array of Josephson junctions, which is manifested in strong variations of the local DOS and broadening of the resistive transitions with the onset of the transition not changing upon increased disorder. Hence, in our films the electronic properties are not governed by the morphology revealed in topography images.

The first hints of a pair-breaking effect in MoC thin films deposited on $Al_2O_3$ substrate date back to the first transport data of Lee and Ketterson [17]. There, the disorder induced superconductor-insulator transition was observed near $R_s \approx 3$ k$\Omega$, far below the quantum resistance $R_Q$, suggesting, that the localization of quasiparticles leading to a renormalization of Coulomb interaction is not the only cause of the suppression of superconductivity. The analysis of our transport data [5,31] leads to a similar conclusion that superconductivity in MoC/$Al_2O_3$ films is suppressed at about 2.5 k$\Omega$.

Our complex conductivity data [12] measured in MoC coplanar waveguide resonators also point to the existence of finite lifetime effects in superconducting MoC thin films. MoN terahertz measurements [11] show that beside the thermally activated quasiparticles another dissipative channel is present. All this points to the fact that the Dynes-superconductors effect is quite often found in a strongly disordered ultrathin films. Further experimental and theoretical investigations are required to clarify the mechanism introducing particular magnetic pair breakers at different film-substrate interfaces.

## V. CONCLUSIONS

In summary, transport and low temperature STM and STS measurements were performed on 3 nm thin MoC films deposited simultaneously on silicon (MoC/Si) and sapphire (MoC/$Al_2O_3$) substrates. Our samples, which differ only in the substrate material, reveal different values of the superconducting transition temperature at unchanged value of the sheet resistance. STS spectroscopy measurements allowed local studies of the superconducting quasiparticle spectra. On both types of samples if taken on flat surfaces these spectra are spatially homogeneous but significantly broadened in comparison with the BCS DOS. Resulting gapless DOS with non-thermal quasi-particles at the Fermi level can be described by the Dynes version of the BCS formula, pointing to the existence of local pair-breaking fields in agreement with the model of *Herman & Hlubina*. We have shown, that locally thinner parts of the films feature narrower superconducting gap and larger smearing parameter $\Gamma$. MoC/$Al_2O_3$ sample with lower transition temperature $T_c$ and superconducting gap $\Delta$ features larger broadening parameter $\Gamma$ than MoC/Si. Since the two samples differ only by the substrate, the source of the pair breaking must be close to the interface. Uncompensated electron orbitals at the substrate/thin-film interface can lead to local magnetic moments and cause pair breaking. Recent studies on $Al_2O_3$ proved that even ferromagnetism can be induced in this way indicating that our MoC/$Al_2O_3$ interface is a strong pair breaker, stronger than Si substrate. Similar pair-breaking effect at the interface might also be operative in many other disordered thin superconducting films showing strong dissipation deep in the superconducting state.

## ACKNOWLEDGEMENTS

We acknowledge helpful conversations with M. Skvortsov, L. Ioffe, R. Hlubina, C. Berthod, T. Cren, and J.G. Rodrigo. This work was supported by the projects APVV-14-0605, VEGA 1/0409/15, VEGA 2/049/16, COST CA16218 Nanocohybri and by the U.S. Steel Košice.

[1] A. A. Abrikosov and L. P. Gor'kov, Zh. Eksp. Teor. Fiz. **39**, 1781 (1960) [Sov. Phys. JETP 12, 1243 (1961)].

[2] K. Maki, Prog.Theor. Phys. (Kyoto) **29**, 333 (1963)

[3] P.G. deGennes, Phys. Kondens. Materie **3**, 79 (1964)

[4] S. Skalski, O. Betbeder-Matibet and P. R. Weiss, Phys. Rev. A **136**, 1500 (1964).

[5] P. Szabó, T. Samuely, V. Hašková, J. Kačmarčík, M. Žemlička, M. Grajcar, J. G. Rodrigo, and P. Samuely, Phys. Rev. B **93**, 014505 (2016).

[6] P. W. Anderson, J. Phys. Chem. Solids **11**, 26 (1959).

[7] R. C. Dynes, V. Narayanamurti, and J. P. Garno, Phys. Rev. Lett. **41**, 1509 (1978).

[8] Y. Noat, V. Cherkez, C. Brun, T. Cren, C. Carbillet, F. Debontridder, K. Ilin, M. Siegel, A. Semenov, H.-W. Hübers, and D. Roditchev, Phys. Rev. B **88**, 014503 (2013)

[9] A. Kamlapure, M. Mondal, M. Chand, A. Mishra, J. Jesudasan, V. Bagwe, L. Benfatto, V. Tripathi and P. Raychaudhuri, Appl. Phys. Lett. **96**, 072509 (2010).

[10] P. C. J. J. Coumou, E. F. C. Driessen, J. Bueno, C. Chapelier, and T. M. Klapwijk, Phys. Rev. B **88**, 180505 (2013).

[11] J. Simmendinger, U. S. Pracht, L. Daschke, T. Proslier, J. A. klug, M. Dressel, and M. Scheffler, Phys. Rev. B **94**, 064506 (2016).

[12] M. Žemlička, P. Neilinger, M. Trgala, M. Rehák, D. Manca, M. Grajcar, P. Szabó, P. Samuely, Š. Gaži, U. Hübner, V. M. Vinokur, E. Il'ichev, Phys. Rev. B **92**, 224506 (2015.)

[13] František Herman and Richard Hlubina, Phys. Rev. B **94**, 144508 (2016).

[14] M. Trgala, M. Žemlička, P. Neilinger, M. Rehák, M. Leporis, Š. Gaži, J. Greguš, T. Plecenik, T. Roch, E. Dobročka, M. Grajcar, Appl. Surf. Sci. **312**, 216 (2014).

[15] E.L. Haase, J. Low. Temp. Phys. **69**, 246 (1987).

[16] M. Morita, T. Ohmi, E. Hasegawa, M. Kawakami and M. Ohwada, J. App. Phys. **68,** 1272 (1990)

[17] S. J. Lee and J. B. Ketterson, Phys. Rev. Lett. **64**, 3078 (1990).

[18] L. J. van der Pauw, Philips Res. Repts. **13**, 1-9 (1958), *ibid.* Philips Tech. Rev. 20, 220-224 (1958).

[19] B.L. Altshuler, and A.G. Aronov, Solid State Commun. **36**, 115 (1979).

[20] B. L. Altshuler, A.G. Aronov and P.A. Lee, Phys. Rev. Lett. **44**, 1288 (1980).

[21] R. J. Hamers, R. M. Tromp, and J. E. Demuth, Phys. Rev. Lett. **56**, 1972 (1986).

[22] V. F. Gantmakher, and V. T. Dolgopolov, Phys. – Usp. **53**, 1 (2010).

[23] Yu. N. Ovchinnikov, JETP **37**, 366 (1973).

[24] H. Takada and Y. Kuroda, Solid State Commun. **41**, 643 (1982).

[25] A. M. Finkel'stein, JETP Lett., **45**, 46 (1987).

[26] T. Proslier, J. F. Zasadzinski, L. Cooley, C. Antoine, J. Moore, J. Norem, M. Pellin and K. E. Gray, Appl. Phys. Lett. **92**, 212505 (2008)

[27] Steven C. Erwin and F. J. Himpsel, Nature Communications 1:58 (2010).

[28] Gregory Kopnov, Zeev Vager, and Ron Naaman, Adv. Mater., **19**, 925–928 (2007)

[29] Q. Li,1, J. Xu, J. Liu, H. Du, and B.Ye, J. Appl. Phys. **117**, 233904 (2015).

[30] A. Sundaresan, C.N.R. Rao, Solid State Commun. **149,** 1197 (2009).

[31] M. Kopčík et al. to be published

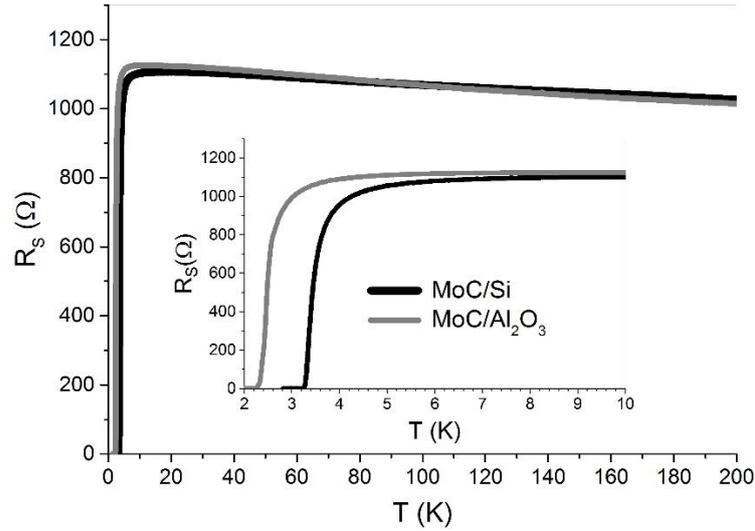

**FIG. 1**. Temperature dependence of sheet resistance $R_s$ of 3nm MoC films sputtered on silicon (black line) and sapphire (gray line). The inset shows $R_s(T)$ near $T_c$.

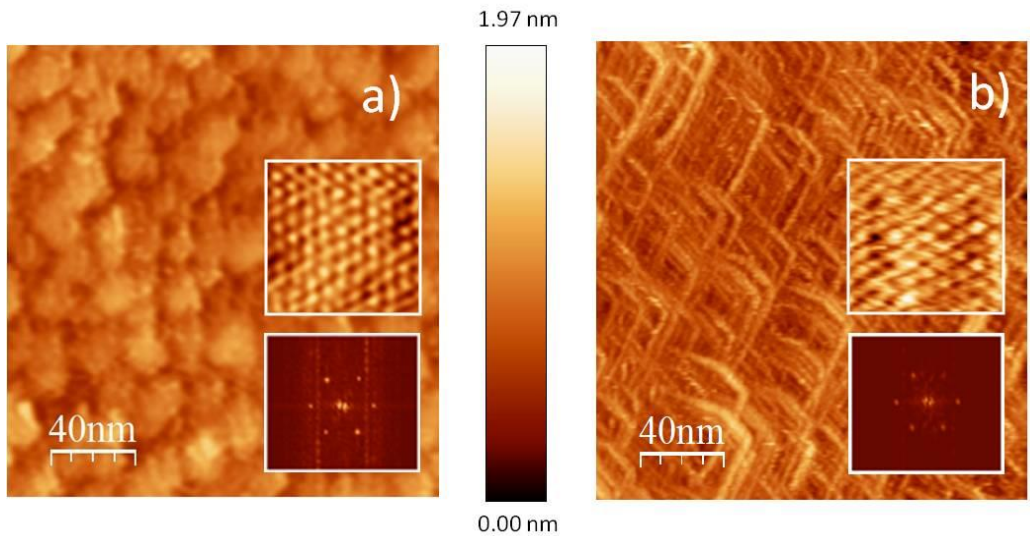

**FIG. 2**. STM topography of 200 nm x 200 nm surfaces of 3nm MoC films sputtered on sapphire (MoC/Al$_2$O$_3$) a) and silicon substrates (MoC/Si) b). The color scale in the middle represents surface corrugation (the insets are not to scale). The insets: upper - 3 nm x 3 nm images of the atomic lattice, bottom - the corresponding Fourrier transforms. The Fourier patterns show $a = a_0\sqrt{2} = 0.6$ nm positions for both lattices, responsible for the 111 plane of a B1 lattice with a lattice constant $a_0 = 0.42$ nm.

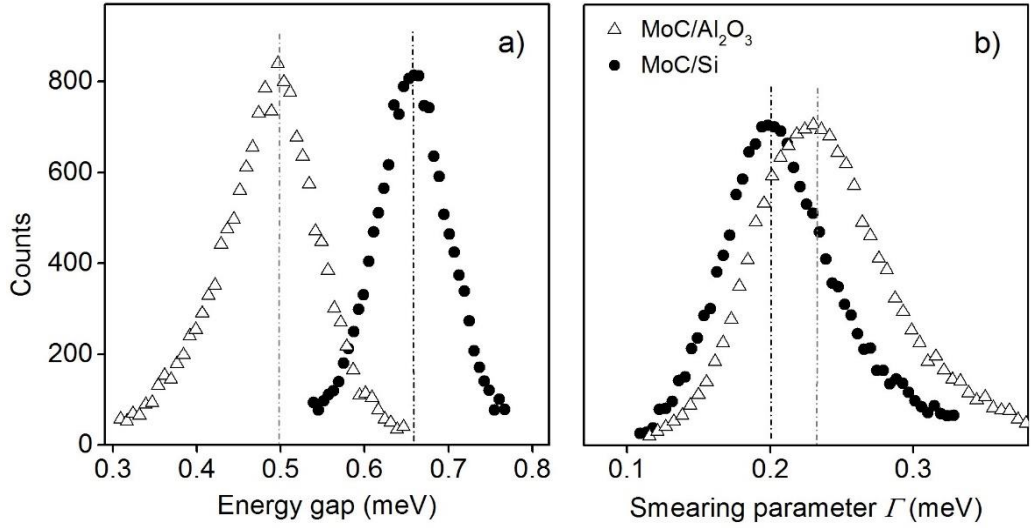

**FIG. 3**. Statistical distribution of a) the superconducting energy gap $\Delta$ and b) spectral smearing parameter $\Gamma$ obtained from the Dynes fit of 128x128 tunneling spectra measured in a 100x100nm$^2$ area of MoC/Al$_2$O$_3$ (open triangle symbols) and MoC/Si (black circles) thin films at $T = 450$ mK.

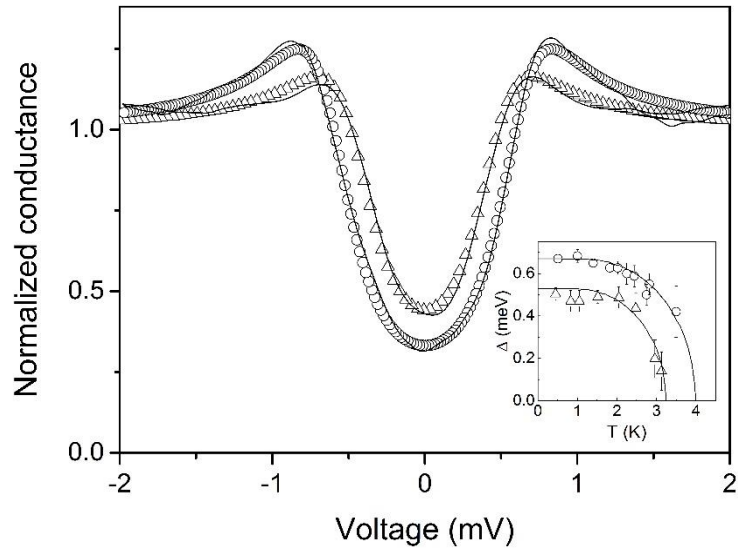

**FIG. 4**. Typical tunneling spectra of 3 nm thin MoC films measured at $T = 450$ mK. The symbols are the Dynes fits (MoC/Si – open circles, MoC/Al$_2$O$_3$ – open triangles). Inset: temperature dependencies of the energy gap of MoC/Si (open circles) and MoC/Al$_2$O$_3$ (open triangles) determined from fitting to the Dynes model. Solid lines show predictions of the BCS theory.

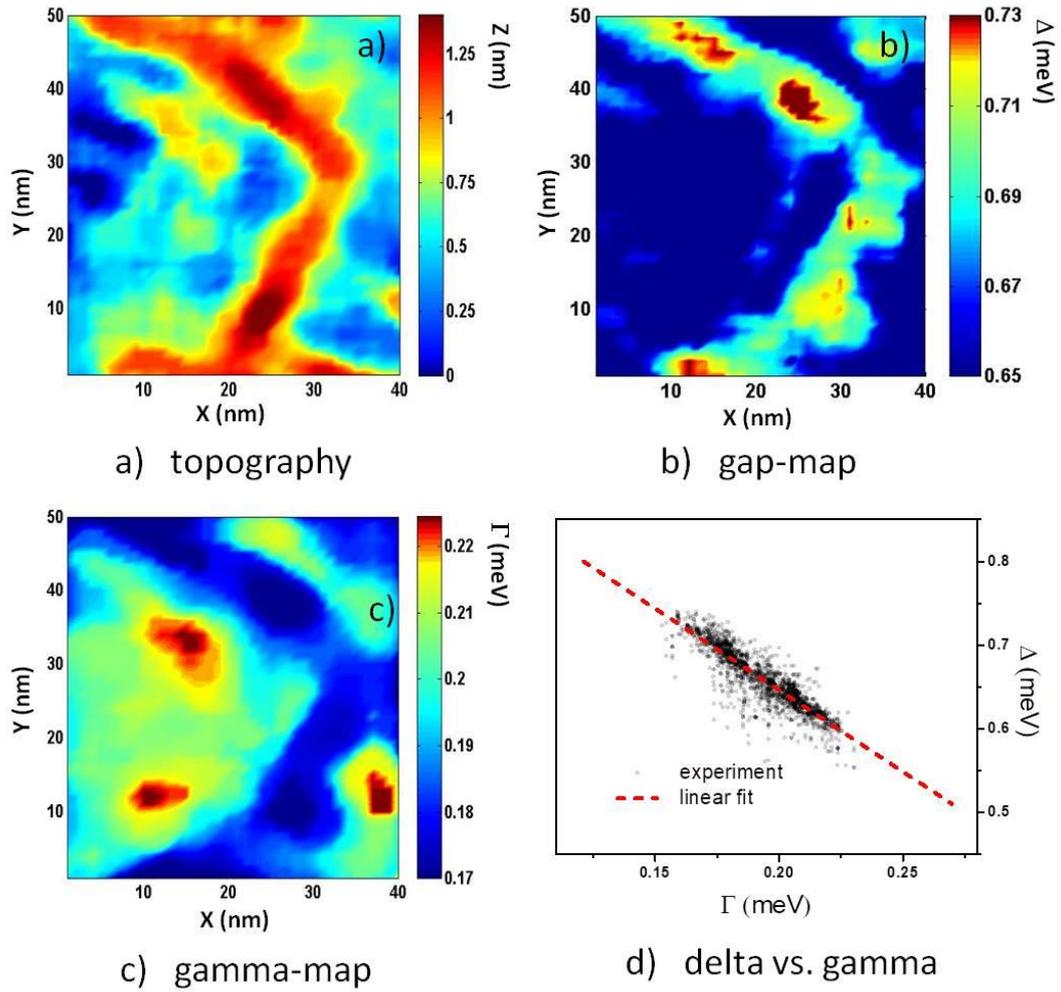

**FIG. 5**. STM topography of a 40 nm x 50 nm surface of MoC/Si thin film – color scale in nm a). Gap-map b) and gamma-map c) of the same area determined from the Dynes formula fits of tunneling spectra measured in a grid of 128x128 points at $T$ = 450 mK. d) Energy gap $\Delta$ as a function of spectral smearing $\Gamma$ constructed from data in b) and c).